\documentclass[%
 reprint,
 amsmath,amssymb,
 aps,
]{revtex4-2}

\usepackage{float}
\usepackage{makecell}
\usepackage{lineno}
\usepackage{graphicx}
\usepackage{dcolumn}
\usepackage{bm}
\usepackage{hyperref}

\begin{document}

\title{Experimental observation and manipulation of optical tornado waves}
\author{Lai Chen}
\affiliation{Department of Physics, Zhejiang University, Hangzhou 310027, China}
\author{Li-Gang Wang}
\email{lgwang@zju.edu.cn}
\affiliation{Department of Physics, Zhejiang University, Hangzhou 310027, China}

\begin{abstract}
We report experimental generation and manipulation of optical tornado waves (ToWs). By controlling the self-focusing length, total angular momentum, and foci deviation of ToWs, the propagation properties of optical ToWs, especially their angular velocity of the main intensity lobes, can be manipulated. We achieve controlling the accumulated rotation angle of the intensity lobes from 0 to 1100 degrees. Also, we confirm that ToWs get the highest angular velocity around the foci coincide situation. Our experimental results are in good agreement with numerical results.
\end{abstract}

\date{\today}
\maketitle

The study of structured light fields is of great interest in optics \cite{padgett2017,durnin1987,efremidis2019,tamburini2006,gibson2004,padgett2011}. The structured light fields, such as vortex beams \cite{padgett2017}, Bessel beams \cite{durnin1987} and Airy beams \cite{efremidis2019}, always exhibit novel dynamic properties. The vortex beams with phase singularities carry orbital angular momentum (OAM). It is proved that the optical beams with vortex can overcome the Rayleigh criterion limit \cite{tamburini2006} and OAM also provides a new degree for communication \cite{gibson2004}. The transverse orbital-associated energy flow makes vortex beams useful in optical twist \cite{padgett2011}.

In the same context, Airy beams exhibit self-accelerating \cite{siviloglou2007} and self-healing properties \cite{broky2008}. Circular Airy beams (CABs) were firstly proposed by Efremidis \textit{et al.} \cite{efremidis2010} and their propagation properties of CABs were widely investigated. As a kind of self-accelerating beams, CABs can self-focus during the propagation and have a strong conical energy flow and intensity gradient near the focus. Thus, CABs have be applied in optical manipulation \cite{zhang2011} and non-linear intense light bullets \cite{panagiotopoulos2013}.

The light fields superimposed by two vortex beams with opposite OAM exhibit angular self-accelerating property during propagation \cite{webster2017,schulze2015,brimis2020}. These components of the superimposed light fields include Laguerre-Gaussian modes \cite{webster2017}, vortex Bessel beams \cite{schulze2015} and circular vortex Airy beams (CVABs) \cite{brimis2020}. Recently, a new kind of light fields superimposed by several CVABs, so-called tornado waves (ToWs), was proposed \cite{brimis2020,mansour2021}, whose main intensity lobes have a very high angular velocity due to the self-focusing property of CVABs. The predicted angular velocity of ToWs \cite{brimis2020} is larger by orders of magnitude than those superimposed light fields using Laguerre-Gaussian modes \cite{webster2017} and Bessel modes \cite{schulze2015}. This novel property provides the potential applications in optical manipulation and laser writing. However, to the best of our knowledge, the detailed method about how to manipulate the transmission properties of ToWs has never been addressed before, which is very important for practical applications.

In this Letter, we experimentally generate optical ToWs and show the influence of the following three physical quantities on ToWs: (1) the self-focusing length, (2) the total angular momentum of ToWs and (3) the deviation of the two foci. Following our proposed method, one can easily manipulate the angular velocity of ToWs. Our experimental and numerical results show that the accumulated rotation angle $\Theta(z)$ of the high intensity lobes can be adjusted from 0 to 1100 degrees.

Let us first briefly introduce optical ToWs. The ToWs superimposed by two CVABs can be described at $z=0$ by \cite{brimis2020}
\begin{align}
  E(r,\varphi,z=0)&=F_1(r)\exp(il_1\varphi)+F_2(r)\exp(il_2\varphi),\\
  F_n(r)&={\rm Ai}\left(\frac{r_n-r}{w_n}\right)\exp\left[\frac{a\left(r_n-r\right)}{w_n}\right],
\end{align}
where ${\rm Ai}(\cdot)$ denotes the Airy function, $r$ and $\varphi$ are the radial and azimuthal coordinates, respectively, $a\ll 1$ is the decay factor, $r_n$, $w_n$ and $l_n$ are the radius of the main ring, the scaling factor, and the topological charge of the $n$-th CVAB, respectively. The self-focusing length of the $n$-th CVAB in free space is given by  $f_{{\rm Ai},n} = 2k\sqrt{w_n^3(r_n + w_n)}$ \cite{papazoglou2011}. The propagation evolution of ToWs in free space can be calculated by Collins formula \cite{collins1970}:
\begin{align}
  E(r,\varphi,z)=&\sum_{n=1}^2 -\frac{k}{i^{l_n-1}z}\exp\left\{i\left[l_n\varphi+k\left(z+\frac{r^2}{2z}\right)\right]\right\}\nonumber\\
  &\times\int_0^{+\infty} F_n(r')\exp\left(\frac{ikr'^2}{2z}\right)J_{l_n}\left(\frac{krr'}{z}\right)r'dr',
\end{align}
where $J_{l_n}$ is the first-kind Bessel function of the $l_n$-th order. The angular momentum density $\vec{j}$ is related to the Poynting vector $\vec{S}$ \cite{allen2000}, and the total angular momentum $\vec{L}$ is the integral of the angular momentum density $\vec{j}$. For a linear-polarized light, the transverse Poynting vector $\vec{S}_{\perp}$ can be expressed as $\vec{S}_{\perp}=(i/2\omega\mu_0)[E\nabla E^*-E^*\nabla E]$, where $\nabla=\hat{r}\partial_r+\hat{\varphi}\partial_\varphi/r$, $\omega=2\pi/\lambda$ is the angular frequency of the light and $\mu_0$ is the permeability of vacuum. Thus, the angular momentum $L$ of ToWs per photon can be express as:
\begin{align}\label{Jz}
  L = \hbar\frac{\sum_{n=1}^2 l_n\int F_n^2(r)rdrd\varphi}{\sum_{n=1}^2 \int F_n^2(r)rdrd\varphi},
\end{align}
where the integral in Eq.(\ref{Jz}) is the total energy of the CVAB. Under the approximation $w_i/r_i\ll1$ and $a\ll 1$, Eq.(\ref{Jz}) can be expressed as:
\begin{align}\label{eq5}
  L=\hbar\frac{\sum_{n=1}^2 l_n\left(r_nw_n/2+w_n^2/8a\right)}{\sum_{n=1}^2\left(r_nw_n/2+w_n^2/8a\right)}.
\end{align}
It is clear that the total angular momentum is realted to the total energy and topological charge of the each CVAB component. Therefore, the angular momentum is not necessarily equal to zero, even if the sum of topological charges is zero.

\begin{figure}[b]
  \centering
  \includegraphics[width=1\linewidth]{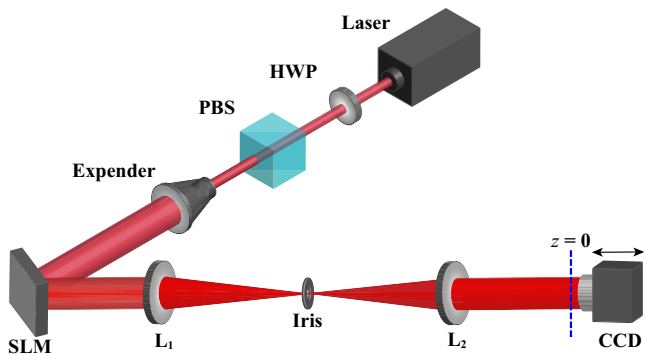}
  \caption{Experimental setup for generating ToWs. The focal plane of the lens $L_2$ is the initial plane $z=0$. Other natations are: HWP, half-wave plate; PBS, polarized beam splitter; SLM, spatial light modulator.}
  \label{fig1}
\end{figure}

Figure \ref{fig1} shows the experimental setup for generating ToWs. A linearly polarized He-Ne laser ($\lambda=633$ nm) passes through the half-wave plate and polarized beam splitter to control the incident light intensity. Then the expended laser incidents on a phase-only spatial light modulator (HoloeyePLUTO-2-VIS-056) with $1920\times 1080$ pixels, where the computer-generated phase-diagram is loaded priorly \cite{davis1999}. In all situations, we take $a=0.05$ in experiments and numericals simulations. Then a $4-f$ system with an iris is used to  select the first-order diffraction light, and here the focal length of the two lenses $L_1$ and $L_2$ are $f_1=f_2=500$ mm. Finally, a ToW is generated at the focal plane of lens $L_2$ as the initial plane $z=0$ and a movable CCD records the intensity distribution at different $z$.

\begin{figure}[t]
  \centering
  \includegraphics[width=\linewidth]{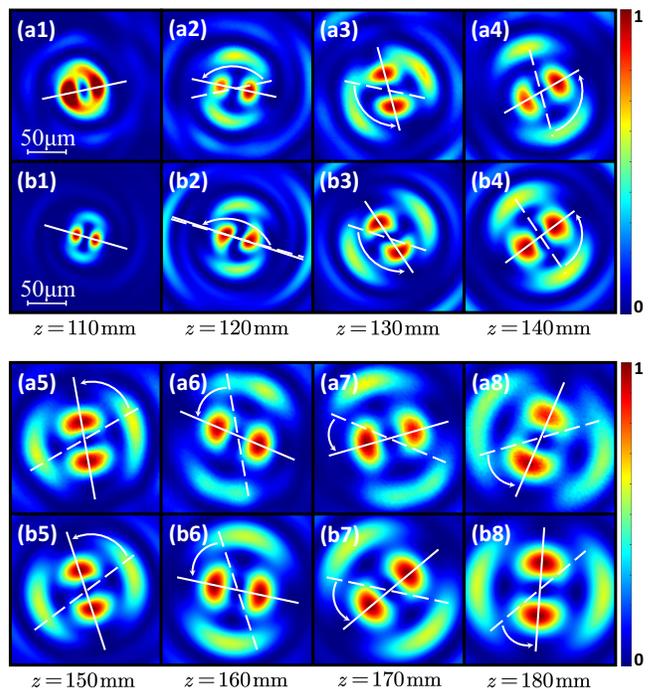}
  \caption{(a) Experimental and (b) numerical results of the intensity lobes with $r_1=1$ mm, $w_1=0.031$ mm, $r_2=0.5$ mm, $w_2=0.0385$ mm. The white solid and dashed lines represent the azimuthal position of the high intensity lobes in the current picture and previous one picture, respectively}
  \label{fig2}
\end{figure}

Figure \ref{fig2} shows the experimental and numerical intensity distributions of a ToW at different propagation distance $z$. In order to show the rotation progress of the ToW intuitively, we use white solid and dashed lines to indicate the azimuthal position of the high intensity lobes in the current and previous positions, respectively. Note that there is roughly a constant longitudinal distance error $\Delta z\approx 3$ mm between the theory and our experiment, i.e., the distance $z$ in Fig.\ref{fig2} is the theoretical distance and the corresponding experimental measured distance is $z+\Delta z$. After self-focusing, the high intensity lobes begin to rotate rapidly and the direction of rotation is anticlockwise in this example. At the beginning, the rotation angle of the ToW reaches nearly 180 degrees within 10 mm in Figs.\ref{fig2}(a1)-\ref{fig2}(a2) and \ref{fig2}(b1)-\ref{fig2}(b2). The rotation velocity decreases rapidly as propagation, and it drops to nearly 50 degrees within 10 mm in Figs.\ref{fig2}(a6)-\ref{fig2}(a8) and \ref{fig2}(b6)-\ref{fig2}(b8). The experimental data are in good agreement with the theoretical prediction. Due to the self-focusing property, the rotation velocity of the ToW is much higher than the superimposed light field using Laguerre-Gaussian \cite{webster2017} and Bessel \cite{schulze2015} modes. Our resluts is also nearly 4 times the reported experimental results \cite{mansour2021}.

\begin{figure}[t]
  \centering
  \includegraphics[width=1\linewidth]{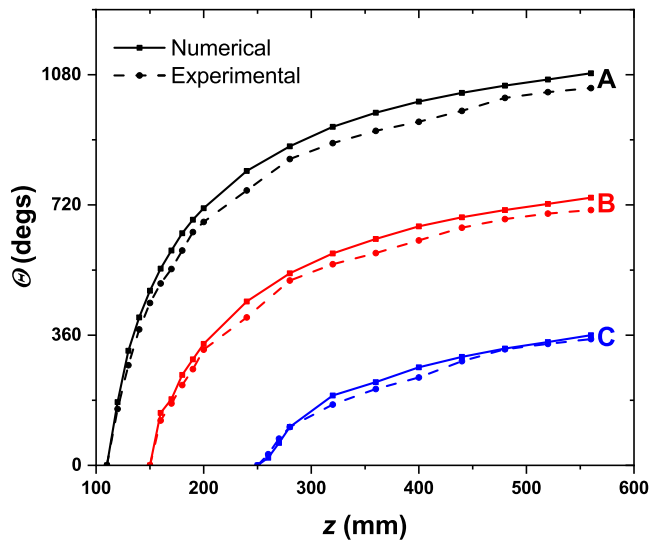}
  \caption{Experimental (dashed curves) and numerical (solid curves) results of rotation angle $\Theta(z)$ as a function of propagation distance $z$, and the corresponding parameters and the maximally accumulated $\Theta(z)_{\max}$ are lised in Table \ref{tab1}.}
  \label{fig3}
\end{figure}

As in Ref.\cite{brimis2020}, according to self-focusing lengths of the two CVAB components, optical ToWs can be divided into two types: (1) foci coincide ($f_{\rm Ai,1}= f_{\rm Ai,2}=f_{\rm Ai}$) and (2) foci partially overlap ($f_{\rm Ai,1}\ne f_{\rm Ai,2}$). In the first case of complete foci overlap, we will discuss the influence of the self-focusing length $f_{{\rm Ai}}$ and the angular momentum $L$ on the angular velocity of the intensity lobes of the ToW. Figure \ref{fig3} shows the experimental and numerical results of the accumulated rotation angle $\Theta(z)$ of the high intensity lobes as a function of propagation distance $z$. The corresponding parameters of the curves A-C are shown in Table \ref{tab1}, which corresponds to the increasing self-focusing length. Here we start recording after the focus. It shows that the rotation angle $\Theta(z)$ increases rapidly after the focus. The increasing rate of $\Theta(z)$ decreases as $z$ increases and finally $\Theta(z)$ gradually tends to be a stead value. Meanwhile, as $f_{\rm Ai}$ increases, the maximally accumulated $\Theta(z)$ becomes small and the rotational angular velocity $\dot{\Theta}(z)$ also decreases. In Fig.\ref{fig3}, the maximal rotation angle of the black curve reaches more than 1080 degrees, which is nearly 3 times the blue curve (only nearly 360 degrees). We can infer at once that it is easy to get a high angular velocity with a small self-focusing length. In principle, it is practicable to get a higher angular velocity by reducing the self-focusing length further. Practically, it requires a higher phase-modulation accuracy in experiments, because the CVAB's ring width will reduce further in this situation.
\begin{table}[t]
\caption{\label{tab1} ToW parameters in Fig.\ref{fig3}}
\begin{ruledtabular}
\begin{tabular}{ccccccc}
 & \makecell[c]{$r_1$\\ (mm)} & \makecell[c]{$w_1$\\ (mm)} & \makecell[c]{$r_2$\\ (mm)} & \makecell[c]{$w_2$\\ (mm)} & \makecell[c]{$f_{\rm Ai}$\\(mm)} & \makecell[c]{$\Theta(z)_{\max}$\\(degs)}\\
\hline
A & 1.00 & 0.0310 & 0.50 & 0.0385 & 110 & 1084\\
B & 1.00 & 0.0380 & 0.50 & 0.0471 & 150 & 740\\
C & 1.00 & 0.0532 & 0.50 & 0.0655 & 250 & 360\\
\end{tabular}
\end{ruledtabular}
\end{table}

\begin{figure}[htbp]
  \centering
  \includegraphics[width=1.0\linewidth]{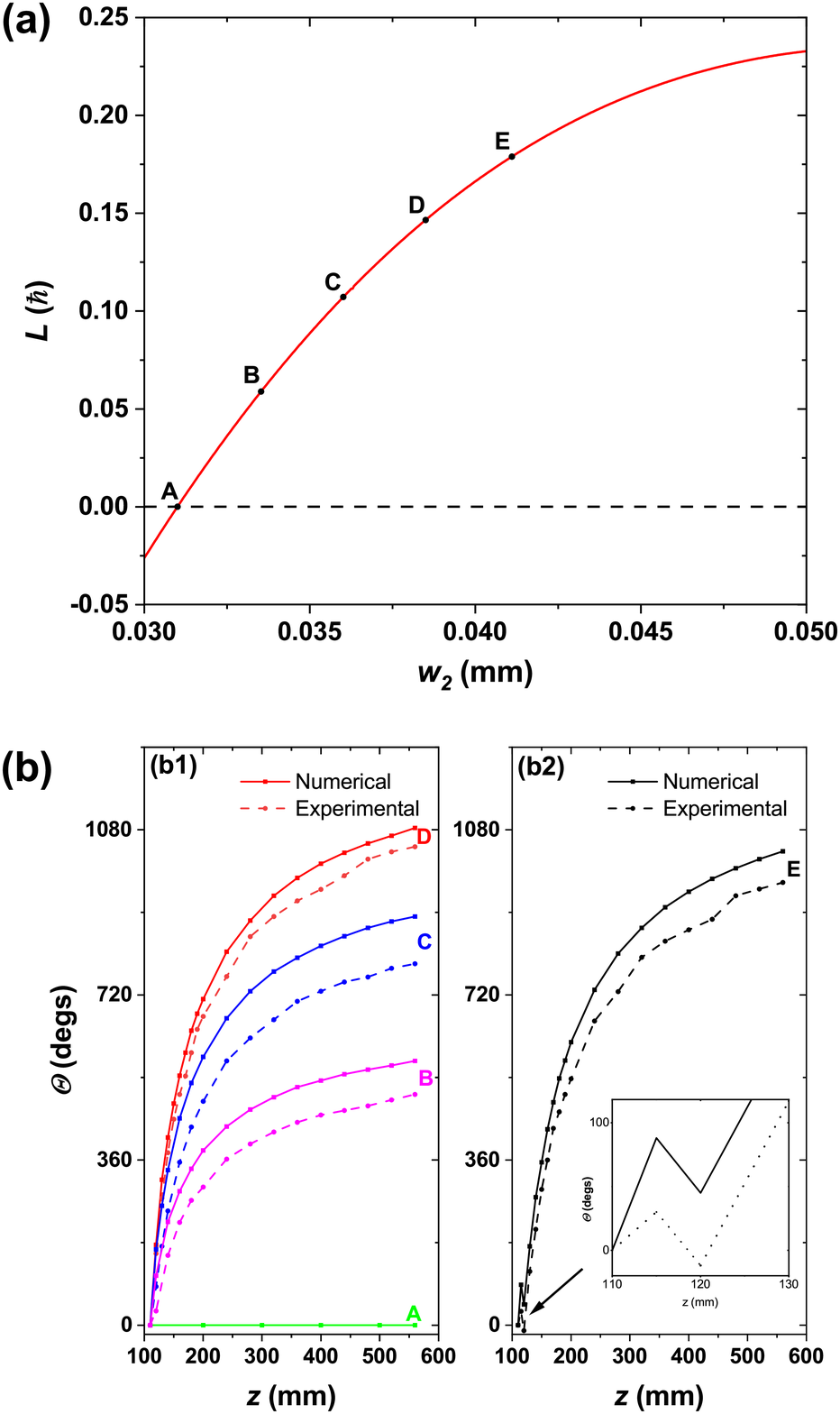}
  \caption{(a) The value of $L$ as a function of $w_2$ with the fixed values of $r_1=$ 1.00 mm and $w_1=$ 0.0310 mm. The zero point of $L(w_2)$ is $w_2=0.0310$ mm. The 5 sets of parameters are lised in Table \ref{tab2}. (b) The accumulated rotation angle $\Theta(z)$ of the high intensity lobes for five cases A-E in (a).}
  \label{fig4}
\end{figure} 
\begin{table}[t]
\caption{\label{tab2} ToW parameters in Fig.\ref{fig4}}
\begin{ruledtabular}
\begin{tabular}{cccccc}
  & $r_1$ (mm) & $w_1$ (mm) & $r_2$ (mm) & $w_2$ (mm) & $\Theta$ (degs)\\
  \hline
  A & 1.00 & 0.0310 & 1.00 & 0.0310 & 0\\
  B & 1.00 & 0.0310 & 0.78 & 0.0335 & 576\\
  C & 1.00 & 0.0310 & 0.62 & 0.0360 & 891\\
  D & 1.00 & 0.0310 & 0.50 & 0.0385 & 1084\\
  E & 1.00 & 0.0310 & 0.40 & 0.0411 & 1033\\
\end{tabular}
\end{ruledtabular}
\end{table}

Next, we consider the influence of the total angular momentum $L$ of ToW on the angular velocity of the high intensity lobes. In the foci overlap situation, the parameters of the ToW satisfy the constraint equation $r_2=(r_1w_1^3+w_1^4-w_2^4)/w_2^3$, and here we set $r_1=1.00$ mm, $w_1=0.0310$ mm. Thus, the angular momentum of ToW is now a function of $w_2$, see Eq.(\ref{eq5}). Next, we will adjust the angular momentum of the ToW by changing $w_2$ to explore its influence on the rotational angular velocity of the high intensity lobes. Fig.\ref{fig4}(a) shows the value of $L$ as a function of $w_2$, and here we choose five cases A-E on the curve and all the parameters are listed in Table \ref{tab2}. Meanwhile, the accumulated rotation angle $\Theta(z)$ of the high intensity lobes under these five cases are plotted in Fig.\ref{fig4}(b). When $L=0$, the high intensity does not rotate as propagating, see the green curve A in Fig.\ref{fig4}(b1). As $L$ increases, it demonstrates that the accumulated rotation angle $\Theta(z)$ and the angular velocity $\dot{\Theta}(z)$ increase, see the curves B-D in Fig.\ref{fig4}(b1). Note that the influence of the angular momentum $L$ on the angular velocity is not linear. From the curve A to the curve B, the value of $\Theta(z)$ has increased about 576 degrees, while from the curve C to the curve D, the value of $\Theta(z)$ has only increased about 200 degrees, see Fig.\ref{fig4}(b1). However, if $L$ continues to increase, the high intensity lobes will begin to rotate in the reverse direction within a short distance, see the black curve E in Fig.\ref{fig4}(b2) around $z\approx 120$ mm. In this situation, the accumulated rotation angle $\Theta(z)$ of curve E is near 1033 degs, which decreases compared with the curve D.

\begin{figure}[t]
  \centering
  \includegraphics[width=\linewidth]{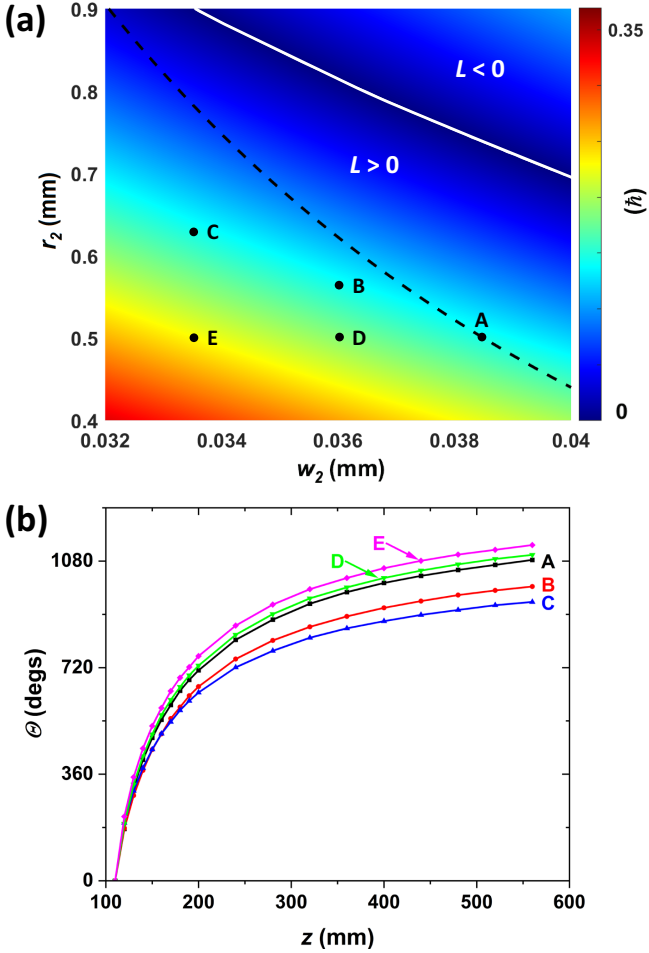}
  \caption{(a) The 2D distribution of $|L(w_2,r_2)|$, here $r_1=$ 1.00 mm, $w_1=$ 0.0310 mm. The 5 sets of parameters are lised in Table \ref{tab3}. $L=0$ is highlighted with the thick solid white curve, and the dashed black curve corresponds to the foci coincide situation. (b) The accumulated rotation angle $\Theta(z)$ of the high intensity lobes for five cases A, B, C, D, and E shown as the bold black points in (a).}
  \label{fig5}
\end{figure}
\begin{table}[t]
\caption{\label{tab3} ToW parameters in Fig.\ref{fig5}}
\begin{ruledtabular}
  \begin{tabular}{ccccccccc}

    & & $r_2$ (mm) & & $w_2$ (mm) & & $\tilde{f}$ & & $\Theta$ (degs)\\
    \hline
    A & & 0.500 & & 0.0386 & & 0     & & 1084\\
    B & & 0.561 & & 0.0360 & & 0.023 & & 994\\
    C & & 0.624 & & 0.0337 & & 0.058 & & 942\\
    D & & 0.500 & & 0.0360 & & 0.053 & & 1101 \\
    E & & 0.500 & & 0.0337 & & 0.100 & & 1134 \\
\end{tabular}
\end{ruledtabular}
\end{table}
Now, let us consider the more general situation, i.e., the foci partially overlap situation.  Here, we will explore the influence of the foci deviation  and the angular momentum $L$ on the rotational angular velocity of the high intensity lobes of ToWs. In order to measure the deviation of the two foci, we define:
\begin{align}
  \tilde{f}=\frac{|f_{{\rm Ai},1}-f_{{\rm Ai},2}|}{f_{{\rm Ai},1}+f_{{\rm Ai},2}}.
\end{align}
Clearly, we have parameter $\tilde{f}\in[0,1)$, when $\tilde{f}=0$, it represents the two foci overlap situation and when $0<\tilde{f}<1$, it corresponds to the foci partially overlap situation.

Figure \ref{fig5}(a) shows the two-dimensional distribution of angular momentum $L$ as functions of $r_2$ and $w_2$: $|L(r_2,w_2)|$. $L=0$ is highlighted with the thick solid white curve and the direction of the angular momentum is opposite on both side of it. The dashed black curve corresponds to the foci coincide situation when one fixs the values of $r_1$ and $w_1$. The rest area corresponds to the foci partially overlap situation. Here we choose 5 sets of parameters in Fig.\ref{fig5}(a), which are indicated by five bold-black points, and all the parameters are listed in Table \ref{tab3}. Meanwhile, Fig.\ref{fig5}(b) shows the accumulated rotation angle $\Theta(z)$ of the high intensity lobes for five cases A-E in Fig.\ref{fig5}(a). In the case of A, B and C in Fig.\ref{fig5}(a), the total angular momentum $L$ is a constant, while their corresponding foci deviation  $\tilde{f}=$ 0, 0.023 and 0.058, respectively. It shows that the ToW with lower foci deviation demonstrates a higher angular velocity and it reaches the maximal angular velocity in the case of complete foci overlap (i.e., $\tilde{f}=0$, see the black curve A in Fig.\ref{fig5}(b)). However, in the case of A and D in Fig.\ref{fig5}(a), both of the total angular momentum $L$ and the foci deviation $\tilde{f}$ of D are larger than A. Larger total angular momentum  accelerates the rotational angular velocity, while larger foci deviation decreases the rotational angular velocity. Under the simultaneous influence of the total angular momentum and the foci deviation, the rotational angular velocity and the accumulated rotation angle of D is slightly larger than those of A. While the cases of D and E are similar.

In conclusion, we have experimentally realized the optical ToWs and investigated their propagation properties. We have proposed three physical quantities that influence the angular velocity of the intensity lobes of ToWs: the self-focusing length, the total angular momentum, and the deviation $\tilde{f}$ of the two foci. In the case of complete foci overlap, it is practicable to get a higher angular velocity by decreasing the self-focusing length and increasing the angular momentum $L$ of the ToW. In the general situation, i.e., the foci partially overlap situation, the angular velocity will be influenced by the foci deviation and the total angular momentum at the same time. By keeping the angular momentum of ToW as a constant and controlling the foci deviation, we have confirmed that the ToW with lower foci deviation demonstrates a higher angular velocity and it reaches the maxium angular velocity in the case of complete foci overlap. Our results may provide potential applications in optical manipulation, laser writing and beam-propagation technologies.

\begin{acknowledgments}
National Natural Science Foundation of China (NSFC) (grant No. 11974309), and National Key Research and Development Program of China (No. 2017YFA0304202).
\end{acknowledgments}


\begin{thebibliography}{99}

    \bibitem{padgett2017} M. J. Padgett, Opt. Express \textbf{25}, 11265 (2017).

    \bibitem{durnin1987} J. Durnin, J. J. Miceli, and J. H. Eberly, Phys. Rev. Lett. \textbf{58}, 1499 (1987).

    \bibitem{efremidis2019} N. K. Efremidis, Z. Chen, M. Segev, and D. N. Christodoulides, Optica \textbf{6}, 686 (2019).

    \bibitem{tamburini2006} F. Tamburini, G. Anzolin, G. Umbriaco, A. Bianchini, and C. Barbieri, Phys. Rev. Lett. \textbf{97}, 163903 (2006).

    \bibitem{gibson2004} G. Gibson, J. Courtial, M. Padgett, M. Vasnetsov, V. Pasko, S. Barnett, and S. Franke-Arnold, Opt. Express \textbf{12}, 5448 (2004).

    \bibitem{padgett2011} M. Padgett and R. Bowman, Nat. photonics \textbf{5}, 343 (2011).

    \bibitem{siviloglou2007} G. A. Siviloglou and D. N. Christodoulides, Opt. Lett. \textbf{32}, 979 (2007).

    \bibitem{broky2008} J. Broky, G. A. Siviloglou, A. Dogariu, and D. N. Christodoulides, Opt. Express \textbf{16}, 12880 (2008).

    \bibitem{efremidis2010} N. K. Efremidis and D. N. Christodoulides, Opt. Lett. \textbf{35}, 4045 (2010).

    \bibitem{zhang2011} P. Zhang, J. Prakash, Z. Zhang, M. S. Mills, N. K. Efremidis, D. N. Christodoulides, and Z. G. Chen, Opt. Lett. \textbf{36}, 2883 (2011).

    \bibitem{panagiotopoulos2013} P. Panagiotopoulos, D. G. Papazoglou, A. Couairon, and S. Tzortzakis,  Nat. Commun. \textbf{4}, 2622 (2013).

    \bibitem{webster2017} J. Webster, C. Rosales-Guzman, and A. Forbes, Opt. Lett. \textbf{42}, 675 (2017).

    \bibitem{schulze2015} C. Schulze, F. S. Roux, A. Dudley, R. Rop, M. Duparre, and A. Forbes, Phys. Rev. A \textbf{91}, 043821 (2015).

    \bibitem{brimis2020} A. Brimis, K. G. Makris, and D. G. Papazoglou, Opt. Lett. \textbf{45}, 280 (2020).

    \bibitem{mansour2021} D. Mansour, A. Brimis, K. G. Makris, and D. G. Papazoglou, “Generation of tornado waves,” in Conference on Lasers and Electro-Optics, (Optical Society of America, 2021), p. FTh1J.7.

    \bibitem{papazoglou2011} D. G. Papazoglou, N. K. Efremidis, D. N. Christodoulides, and S. Tzortzakis, Opt. Lett. \textbf{36}, 1842 (2011).

    \bibitem{collins1970} S. A. Collins, J. Opt. Soc. Am. \textbf{60}, 1168 (1970).

    \bibitem{allen2000} L. Allen and M. J. Padgett, Opt. Commun. \textbf{184}, 67 (2000).

    \bibitem{davis1999} J. A. Davis, D. M. Cottrell, J. Campos, M. J. Yzuel, and I. Moreno, Appl. Opt. \textbf{38}, 5004 (1999).

\end{thebibliography}
\end{document}